\documentclass[11pt]{article}
\usepackage{graphicx}
\usepackage{siunitx}
\usepackage{lineno}
\usepackage{hyperref}
\usepackage{textcomp}  % for \textsuperscript and \dagger safely
\hypersetup{
    hidelinks
}

\title{Correcting Curvature in Micromirror-Based Spatial Light Modulators with a Microlens Array}

\author{
Munkyu Kang\textsuperscript{1,*},
Elizabeth Murray\textsuperscript{1},
Leyla A. Kabuli\textsuperscript{1},\\
Rikky Muller\textsuperscript{1,†},
and Laura Waller\textsuperscript{1,†}
}

\date{}  % remove date

\makeatletter
\@namedef{r@fig:Introduction}{{1}{}{}{}{}}
\@namedef{r@fig:SLM structure and simulation}{{2}{}{}{}{}}
\@namedef{r@fig:Concept of MLA and simulation}{{3}{}{}{}{}}
\@namedef{r@fig:Experimental setup}{{4}{}{}{}{}}
\@namedef{r@fig:Experimental results}{{5}{}{}{}{}}
\@namedef{r@fig:Active phase control}{{6}{}{}{}{}}
\@namedef{r@fig:mirror}{{S1}{}{}{}{}}
\@namedef{r@fig:transmission}{{S2}{}{}{}{}}
\@namedef{r@fig:rotation}{{S3}{}{}{}{}}
\@namedef{r@fig:focal_length}{{S4}{}{}{}{}}
\@namedef{r@fig:635nm}{{S5}{}{}{}{}}
\@namedef{r@fig:wl_simul}{{S6}{}{}{}{}}
\makeatother

\begin{document}
\maketitle

\noindent
\textsuperscript{1}Department of Electrical Engineering and Computer Sciences, University of California, Berkeley, California 94720, USA\\
\textsuperscript{*}\texttt{munkyu@berkeley.edu}\\
\textsuperscript{†}Equal contribution to this work as co-senior authors.

\begin{abstract}
Computer generated holography requires high-speed spatial light modulators (SLMs) for dynamically patterning light in 3D. Piston-motion micromirror-based SLMs support high-speed ($\geq$ 10 kHz) phase modulation; however, fabricating micromirror arrays with sufficient fill factor necessary for high diffraction efficiency is challenging. In particular, the larger mirrors of high fill factor designs are susceptible to stress-induced curvature that significantly degrades optical performance. In this work, we introduce an optical compensation method using a pitch-matched microlens array (MLA) to focus light onto just the center of each mirror. Our approach thus avoids curvature-induced artifacts and improves optical fill factor to nearly 100\%, independent of the original mechanical fill factor. Through simulations and experiments on a fabricated micromirror array with bowed mirrors, we show that the Pearson correlation coefficient of the imparted phase profile is improved from 0.11 to 0.85 and the brightness of a holographically-generated single spot is enhanced by 8$\times$ with our microlens array in place. Our hybrid optical-electromechanical strategy thus provides a scalable path toward high-speed, high-fidelity wavefront control for applications such as adaptive optics, holographic displays, and optogenetics.
\end{abstract}
%%%%%%%%%%%%%%%%%%%%%%%%%%  body  %%%%%%%%%%%%%%%%%%%%%%%%%%
\section{Introduction}

Computer generated holography (CGH) can pattern light intensity in 3D through pixel-level phase control of coherent light. This feature is fundamental to a wide range of emerging technologies, including virtual and augmented reality (VR/AR)~\cite{maimoneARVR, changVRAR}, 3D fabrication~\cite{3dfabrication}, and holographic optogenetics~\cite{pegardOptogenetics, pegardReview}. A CGH system for generating a desired 3D intensity pattern typically consists of two key components -- the computational algorithm that calculates the appropriate phase pattern and a hardware device, most commonly a spatial light modulator (SLM), that physically modulates the optical wavefront based on this phase pattern. High refresh rates are crucial for applications  where rapid updates to the 3D intensity pattern are necessary, such as real-time near-eye holographic displays~\cite{choi2022time} and dynamic neural photo-stimulation~\cite{pegardOptogenetics}. While substantial progress has been made in developing high-speed CGH algorithms~\cite{Shi2021neural, pegardReview, nathan3Dpointcloud}, the overall system refresh rate remains constrained by the speed of the SLM.

The most widely used commercial SLMs are based on liquid crystal on silicon (LCoS) technology, which were originally designed for 2D displays. LCoS SLMs offer continuous phase modulation and high spatial resolution, making them well-suited for many CGH applications~\cite{LCOS-SLMreview}. Phase modulation is achieved by applying a voltage at each pixel to rotate birefringent liquid crystal molecules, which induces localized changes in the refractive index that modulate the phase at that pixel. Most commercially available LCoS SLMs operate at standard video display rates (60fps)~\cite{holoeyegaea}, which is sufficient for traditional 2D displays -- the most common commercial use case. For applications that require faster response rates, efforts to increase the speed of these devices have used strategies such as active heating/cooling and transient high-voltage drive to reach 500-1400 Hz \cite{pegardReview, activeheatingslm, meadowlarkSLM}.

Micro-electromechanical systems (MEMS)-based devices can achieve much higher refresh rates. For example, Texas Instruments' digital micromirror devices (DMDs) operate at 30 kHz switching speeds~\cite{dmd32Hzandgrayscale}. However, DMDs are not well suited for holography because the tip/tilt micromirror architecture can only display binary amplitude holograms with limited diffraction efficiencies \cite{ketchum2021diffraction}. To enable >10~kHz refresh rates and continuous phase modulation for high-efficiency holographic projection, piston-motion micromiror arrays (MMAs) are a promising technology \cite{TIPLMpaper, Nathanannular}. Piston-motion MMAs achieve analog phase modulation by physically moving micromirrors vertically along the optical axis to modulate the optical path length. To maximize diffraction efficiency, MMAs are designed with a high fill factor, as efficiency is limited to the square of the fill factor\cite{Arrizon1999}. Piston-motion MMAs have recently been adopted in AR/VR applications~\cite{tiPLMVRexample}, and we expect their future use in wavefront shaping, microscopy, and dynamic spectroscopy~\cite{dmd32Hzandgrayscale,photoacoustic30khz,highspectroscopy}. 

However, fabricating high fill factor MMAs presents significant challenges. To achieve high fill factor, mirrors are usually designed to be as large as possible such that the spacing between micromirrors in the array is minimized. These large, tightly-packed micromirrors complicate the release process and require the placement of all support structures and control wiring beneath the active mirror surface, rather than surrounding it. Most critically, stress gradients resulting from the thin-film deposition process can induce significant unwanted mirror curvature in larger mirrors~\cite{curvature_compensation, argon, temperatureanneal}. Curved micromirrors focus the light from each pixel, rather than simply phase shifting it, introducing undesirable wavefront distortion that can severely degrade performance.  While many fabrication techniques have been demonstrated to reduce the curvature of released micro-structures (e.g. compensation layers ~\cite{curvature_compensation}, argon ion machining ~\cite{argon}, high-temperature anneals ~\cite{temperatureanneal}), these approaches add complexity to the fabrication process and must be tailored to individual designs.

\begin{figure}[h]
    \centering
    \includegraphics[width=0.8\textwidth]{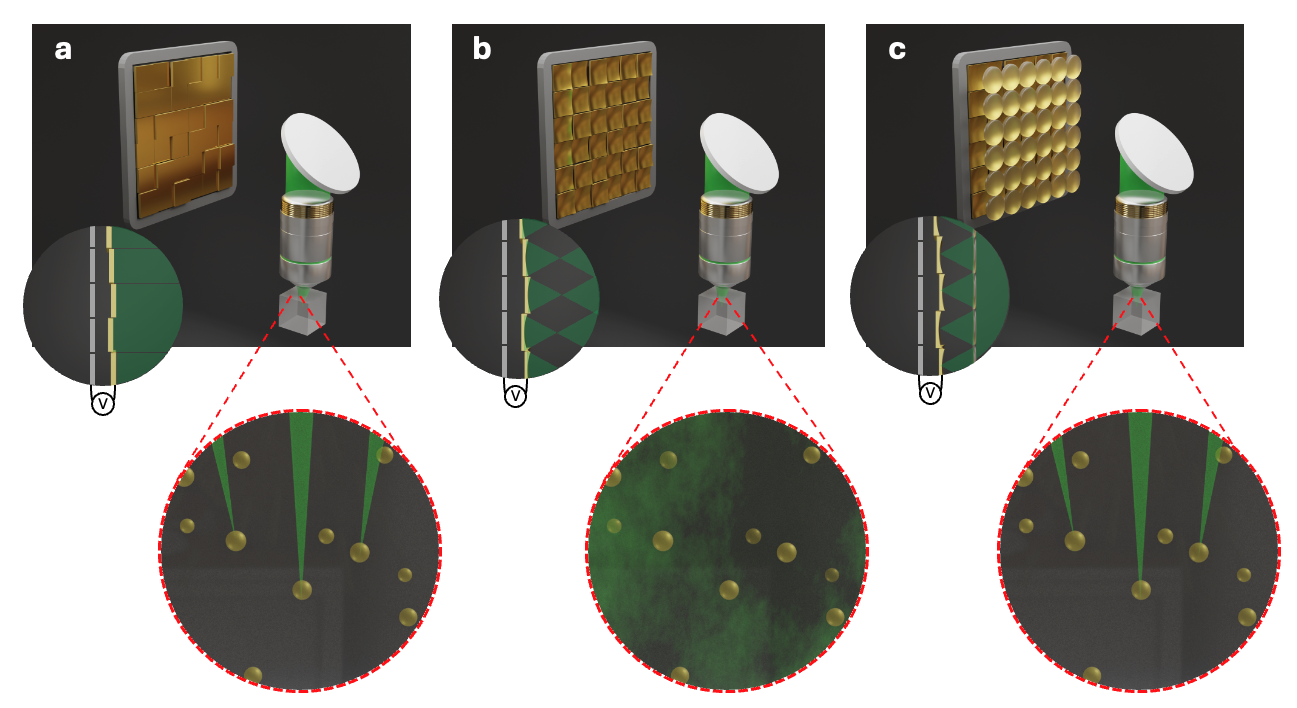}
    \caption{Micromirror compensation technique to improve modulation capability. (a) Ideal micromirrors pattern incident light to form point clouds, which can be used for downstream tasks such as neuromodulation. (b) Due to manufacturing issues, existing micromirror devices have curved surfaces, causing failures in point cloud production. (c) We present a simple and compact optical compensation method: by adding a pixel pitch-matched microlens array in front of the micromirror, we correct for mirror curvature. This results in accurate point cloud formation. }
    \label{fig:Introduction}
\end{figure}

In this work, we propose a method that avoids the problems caused by curvature-induced aberrations by using a microlens array (MLA) to focus the light for each pixel onto the central region of each micromirror (Fig.~\ref{fig:Introduction}c). The MLA pitch matches the SLM's pixel pitch; it is laterally pixel-aligned and axially placed one focal length from the SLM surface. Light goes through the microlenses of the MLA and arrives at the SLM surface in focused spots; thus only the central portion of the micromirror needs to be flat. After reflection, the light travels back in the opposite direction; the focused spots expand and are collimated by the MLA, resulting in a plane wave coming out of the device. By programmable actuation of each pixel's micromirror, we can then pattern the phase at each pixel of the reflected beam without suffering from artifacts due to bowed mirrors. Since the MLA focuses all the incident light onto the active mirror areas, this approach also reduces diffraction artifacts associated with imperfect fill factor. 

MLAs are commonly available, as they are frequently used in other optical setups, including plenoptic imaging~\cite{horisakicompoundImaging, renLightField}, projection photolithography~\cite{wu2002}, beam homogenizers~\cite{KoppBeamshaper} and projection screens~\cite{hediliProjection}. One of the most common uses of MLAs is in commercial image sensors, where they serve a similar purpose as in this work; the MLA concentrates light onto the active part of each pixel, allowing the sensor to have larger optical fill factor (typically up to 3$\times$) without complicated fabrication processes~\cite{CMOSsensors, originalMLACCD}. MLAs were also used to improve fill factor in early piston-motion micromirror arrays for applications such as wavefront correction ~\cite{cowan98MLA, CMOSMLA}. Note that in both image sensors and deformable mirrors for adaptive optics, the device must deal with a relatively large range of angles for incoming light. In computer generated holography (CGH), however, the illuminating light is typically a plane wave; this allows for even smaller proportional active areas in SLM designs for CGH than possible in other applications. MLAs have also been integrated with ferroelectric liquid crystal SLMs to avoid incident light on bumpy surface regions introduced by integrated circuitry, but only for large pixel sizes (\SI{200}{\micro\meter}) and binary modulation~\cite{Chase95VLSIMLA}. While these earlier studies showed that MLAs can improve fill factor and optical efficiency, they were limited to demonstrations on devices with relatively large pixel sizes and for applications other than holography. In contrast, our work targets high-speed piston-motion SLMs for CGH where stress-induced mirror curvature must be corrected to enable accurate, continuous phase modulation. Using high-precision MLA fabrication and MEMS processing, we demonstrate operation with smaller \SI{70}{\micro\meter}-pitch pixels and achieve a 100\% optical fill factor. This approach suggests the possibility of reducing the physical fill factor requirement in future designs, which could simplify MEMS fabrication and improve actuation speed without sacrificing optical performance.

We first quantify the optical phase distortions introduced by mirror bowing in a prototype device and evaluate their impact on holographic 3D point cloud quality. We then provide design principles for MLA compensation and generate an MLA design for the prototype device. Through a combination of optical simulations and experimental validation, we demonstrate that our MLA-based approach effectively restores phase uniformity. 

\section{Impact of Pixel-Level Curvature in Piston-Motion Micromirror SLMs}

\begin{figure}[h]
\centering
\includegraphics[width=1\textwidth]{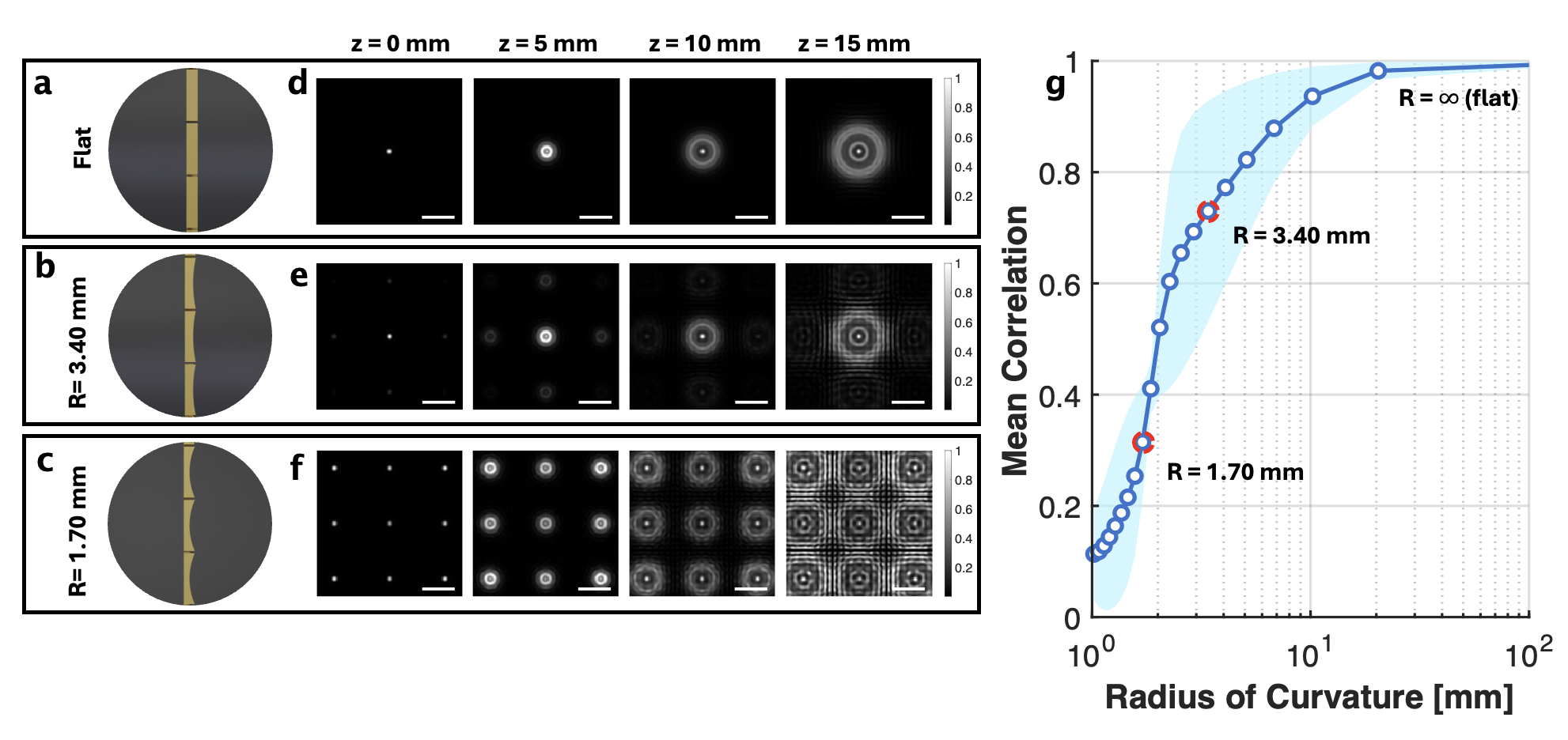}
\caption{Structure of the SLM and simulation of CGH point degradation due to micromirror curvature. (a–c) Schematics of a micromirror with flat (no curvature), mild curvature, and greater curvature, respectively. (d–f) Simulated wave propagation of the focused spot generated by the SLM for each curvature case. From left to right, the propagation distances are $z = 0$, 5, 10, and 15 mm. The colorbar indicates normalized intensity. (g) Mean intensity correlation between curved and flat mirrors as a function of radius of curvature. For each curvature, the propagated intensity patterns were compared with the corresponding flat-mirror reference at the same distance. The correlations across all propagation distances were averaged, and the standard deviation is shown as a shaded light-blue region. Scale bar in (d–f): \SI{200}{\micro\meter}.}
\label{fig:SLM structure and simulation}
\end{figure}

Surface curvature caused by residual stress gradients is a common fabrication issue in surface micromachined MEMS mirrors. The thin films deposited on a mirror surface to achieve high reflectivity (e.g. Au) often exhibit high residual stress, and once sacrificial layers are released to produce suspended structures, the structure bows. Bowing can be particularly severe in designs with large, but thin, mirror bodies, as was the case for a fabricated \SI{70}{\micro\meter}-pitch, 94\%-fill factor  micromirror array used to test the proposed MLA phase flattening technique (Fig. ~\ref{fig:Introduction}b). (Information on the design and fabrication of the micromirror array can be found in the Supplementary Document.) Using digital holographic microscopy (DHM), we find that individual micromirrors in the fabricated array exhibit bowing with a radius of curvature of approximately 1.70 mm. This means that light reflecting off each pixel is actually focused to a spot 0.85 mm from the mirror surface, rather than a flat beam, significantly degrading wavefront fidelity and distorting holographically-generated intensity patterns.

To quantitatively evaluate the impact of mirror curvature on CGH performance, we conducted wave propagation simulations with varying amounts of mirror bowing on a 16$\times$16 micromirror array. Figure~\ref{fig:SLM structure and simulation}a–c illustrates schematic side views of the geometry of micromirrors with varying radius of curvature. Based on these geometries, we applied the angular spectrum method~\cite{goodman2005fourieroptics} to simulate the light propagation through the system, assuming a far-field (Fourier) configuration, which transforms a plane wave to a single point at the detector. (Details of the angular spectrum propagation modeling can be found in the Supplementary Document.) The resulting intensity profiles at different propagation distances (0 mm, 5 mm, 10 mm, and 15 mm) from the desired far-field focus plane are shown in Fig.~\ref{fig:SLM structure and simulation}d–f, for mirrors with a radius of curvature of $\infty$ (flat), 3.40 mm, and 1.70 mm. As the curvature increases, the quality of the beam degrades significantly, with severe aliasing and distortion observed in the 1.70 mm curvature case (which matches the fabricated device), resulting in a grid of spots rather than the intended single focused spot. We computed the intensity correlation between the output of the curved-mirror simulation and the flat-mirror reference, then averaged the correlation across multiple propagation distances to yield a mean correlation, plotted in Fig.~\ref{fig:SLM structure and simulation}g. The mean correlation decreases rapidly with increasing curvature and a radius of curvature greater than approximately 12.25 mm is required to maintain at least 95\% similarity to the flat mirror. In short, even small amounts of mirror curvature can cause substantial degradation in holographic beam quality, emphasizing the need for either improved fabrication or effective correction techniques.

\section{Microlens Array Compensation}
To address the phase distortions introduced by curvature in MEMS-based SLMs, we propose an optical strategy using a MLA. Rather than relying on fabrication-level improvements, our approach employs MLA-based pre-conditioning of the incident wavefront to avoid the curvature-induced artifacts. By placing an MLA one focal length in front of the SLM, the incident light will be focused onto a spot at the central region of each mirror. If the focused spot size is small, there will be minimal distortion due to mirror curvature. If the depth-of-focus of the spot is large relative to the mirror's actuation range ($\sim$300nm in our fabricated device), then the reflected beam will be properly collimated by the MLA after the light reflects and passes through it in the other direction. This not only mitigates curvature-induced aberrations but also improves optical fill factor by redirecting light that would otherwise fall onto non-reflective inter-mirror gaps. 

Choosing the MLA focal length involves several design tradeoffs. A shorter focal length makes the device more compact and offers higher numerical aperture (NA), which means the focused spot will be smaller and thus less affected by the mirror curvature. However, a larger NA also results in increased aberrations and smaller depth-of-focus (Rayleigh range) for the spots, which makes aligning and placing the MLA more difficult. Smaller focal lengths demand more precise alignment between the MLA and the SLM, increasing sensitivity to mechanical drift. Even after carefully aligning the unactuated SLM at the MLA's focal plane, the microlenses will generate a slightly converging beam for actuated pixels; this effect should be negligible in our design, given that the Rayleigh range (>~\SI{100}{\micro\meter}) is much larger than the mirror actuation range (\SI{300}{\nano\meter}). 

To quantify the maximum focal length for ensuring that the focused spot size is smaller than the `flat' region of the mirror, we adopt a flatness criterion requiring the maximum deviation from the tangent plane at the center of the mirror to be less than $\lambda/16$. For a spherical mirror with radius of curvature $R$, the height profile $h(r)$ can be calculated as:

\begin{equation}
h(r) = R- \sqrt{R^2-r^2} \leq \frac{\lambda}{16},
\end{equation}
\noindent where $r$ is the radial distance from the center of the mirror. Therefore, the maximum radius for the area over which the mirror is considered flat is:

\begin{equation}
% r \leq \sqrt{\frac{1}{5}{\left(\lambda R - \frac{\lambda^2}{20}\right)}}.
r \leq \sqrt{\frac{1}{8}(\lambda R - \frac{\lambda^2}{32})}.
\end{equation}

For the fabricated mirrors with $R = 1.53$ mm at wavelength $532$ nm, the radius of the effectively flat region is \SI{10.09}{\micro\meter}, representing approximately $6.52 \%$ of the total mirror area (for our \SI{70}{\micro\meter}-pitch mirrors). %This analysis provides valuable guidance for both using our current SLM with its limitations and for future manufacturing improvements. If we redesign our SLM to only have active micromirrors at these central spots, we could drastically reduce the device fill factor, which should simplify the fabrication process, enable faster mirror actuation, and minimize crosstalk. 

Having established that we need to restrict our beam to approximately \SI{10.09}{\micro\meter} radius (\SI{20.18}{\micro\meter} diameter) to maintain an effectively flat region on our curved mirrors, we now need to determine the appropriate focal length for our microlens array. Each microlens in the array has a diameter of \SI{70}{\micro\meter}, matching the pitch of our mirror array. For a lens with diameter $D$ and focal length $f$, the diameter of the diffraction-limited spot size is given by:

\begin{equation}
s \approx 1.22 \cdot \lambda \cdot \frac{f}{D}.
\end{equation}

Plugging in the diameter of our MLA lenses, $D=$~\SI{70}{\micro\meter}, and the diameter of the spot over which we need the mirror to be effectively flat, $s \leq$ \SI{20.18}{\micro\meter}, we rearrange the equation to solve for the maximum allowable focal length:

\begin{equation}
f_{max} = \frac{s \cdot D}{1.22 \cdot \lambda}= \frac{(20.18 \times 10^{-6}) \cdot (70 \times 10^{-6})}{1.22 \cdot 532 \times 10^{-9}} \approx 2.18 \text{ mm}.
\end{equation}

\noindent We choose a MLA with a focal length of 1.0 mm, providing a spot diameter of \SI{9.27}{\micro\meter}, which is nearly half the size of the maximum flat region of our mirrors. This spot size ensures that the beam remains well within the effectively flat region of our curved mirrors, allowing for accurate phase modulation despite the mirror curvature. Additional simulations showing the dependence of performance on the MLA focal length are provided in the Supplementary Document.

\begin{figure}[h]
    \centering
    \includegraphics[width=0.8\textwidth]{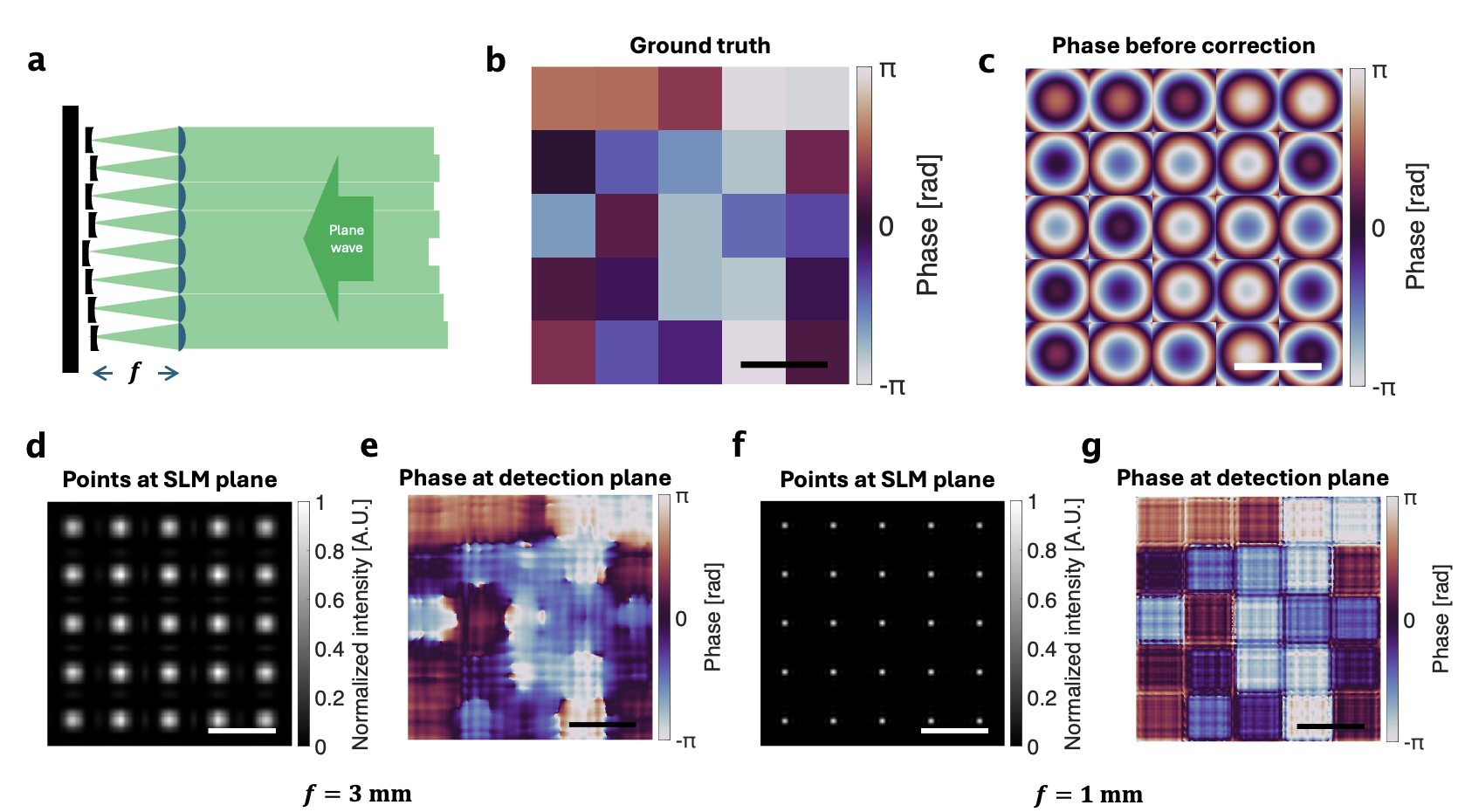}
    \caption{Concept of the MLA and phase flattening simulation. (a) Schematic illustration of phase flattening using a MLA. (b) Initial phase distribution of the micromirrors, each with unique random height displacements. (c) Distorted phase map from the curved mirror SLM. (d) Intensity distribution of the focal spot array at the SLM plane generated by the MLA with a focal length of 3 mm. (e) Phase profile after flattening by the MLA with a 3 mm focal length. (f, g) Same as (d, e), but using an MLA with a 1 mm focal length. The colorbars in (b), (c), (e), and (g) indicate phase in radians. The colorbars in (d) and (f) represent normalized intensity. Scalebars: \SI{100}{\micro\meter}.}
    \label{fig:Concept of MLA and simulation}
\end{figure}

\subsection{Simulations}
To verify the effectiveness of our approach, we conducted beam propagation simulations using a MLA and a curved-mirror SLM (Fig.~\ref{fig:Concept of MLA and simulation}a). Both the MLA and the SLM were modeled as \(5 \times 5\) arrays with \SI{70}{\micro\meter} pitch, where each microlens focuses incoming light onto a corresponding curved micromirror. The micromirrors have random height displacements, each encoding a different phase value, as shown in Fig.~\ref{fig:Concept of MLA and simulation}b. The curvature of the micromirrors produces a bowed phase distortion across each element, shown in Fig.~\ref{fig:Concept of MLA and simulation}c. We show results for two different focal lengths for the MLA, \(f = 3~\mathrm{mm}\) and \(f = 1~\mathrm{mm}\), corresponding to lower (0.012) and higher (0.035) NA, respectively. This allows us to compare how spot size and depth-of-focus affects the reconstruction performance.

A normally-incident plane wave is focused by the MLA to form a \(5 \times 5\) array of focal spots at the SLM plane. The resulting intensity patterns for both focal lengths are shown in Fig.~\ref{fig:Concept of MLA and simulation}d and \ref{fig:Concept of MLA and simulation}f. As expected, the shorter focal length (\(f = 1~\mathrm{mm}\)) produces tighter focal spots due to the higher NA. These reflected wavefronts then propagate back through the MLA and are collimated by the same microlenses. We plot the phase profile at the back focal plane of the MLA. All wave propagation steps were simulated using the angular spectrum method.

Reducing the spot size by using a shorter focal length helps minimize the influence of curved mirror edges. As shown in Fig.~\ref{fig:Concept of MLA and simulation}e and \ref{fig:Concept of MLA and simulation}g, when the focal spot is large (\(f = 3~\mathrm{mm}\)), the light reflects off a wider area of each mirror, causing the intended phase values to become blurred or distorted. In contrast, with a shorter focal length (\(f = 1~\mathrm{mm}\)), each mirror’s phase value is preserved, demonstrating improved phase modulation fidelity. We compared the phase map before correction with that after using the \(f = 1~\mathrm{mm}\) MLA by computing the Pearson correlation coefficient (PCC) with respect to the ground truth. The PCC improved from 0.11 to 0.85, demonstrating a significant enhancement in phase fidelity. Our choice of \(f = 1~\mathrm{mm}\) reflects a balance between achieving sufficient optical confinement and maintaining mechanical robustness. 

\section{Experimental Results}
\begin{figure}[h]
    \centering
    \includegraphics[width=0.8\textwidth]{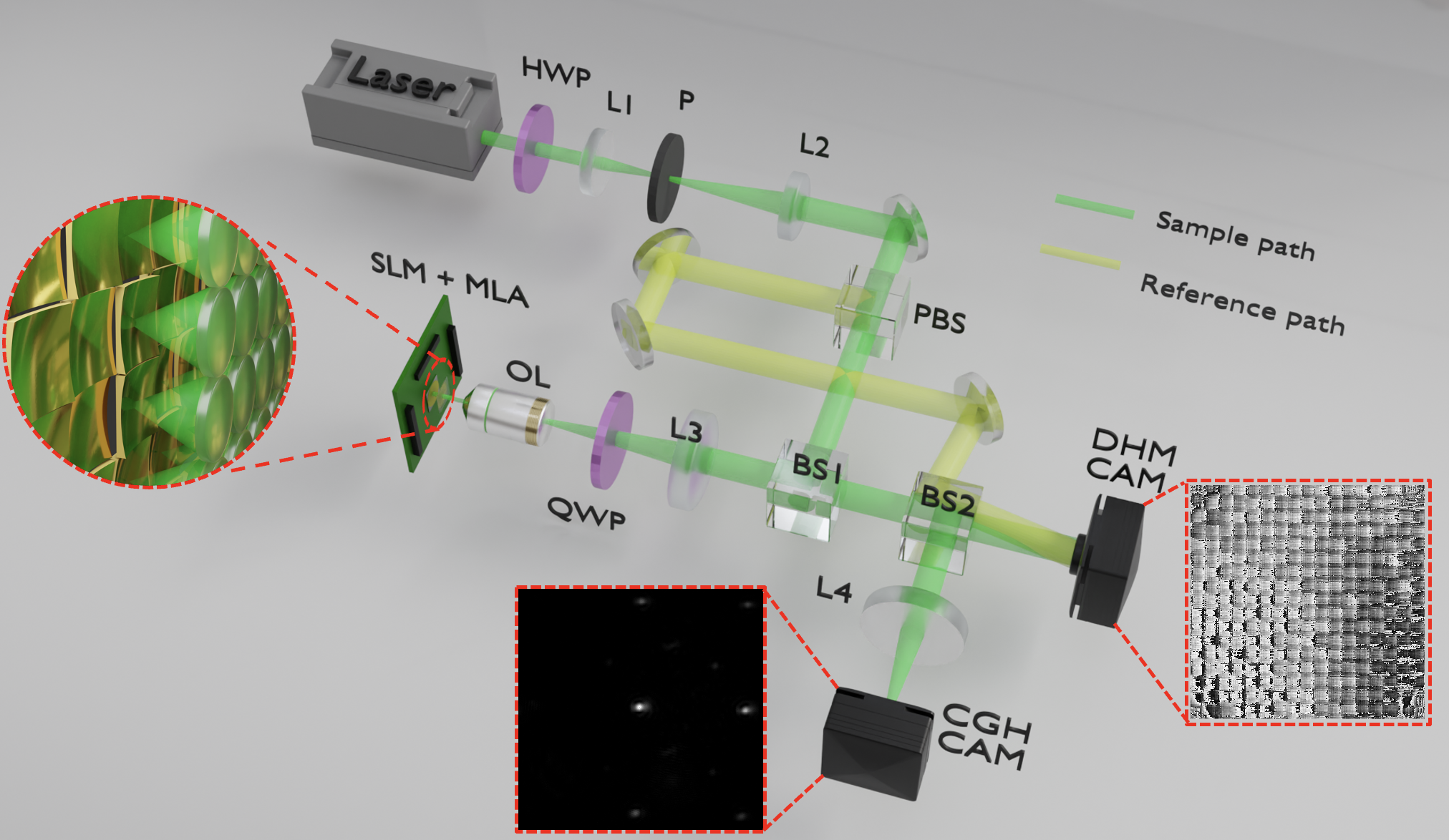}
    \caption{Diagram of the DHM setup combined with the CGH. HWP: half-wave plate, L1-4: lenses, P: pinhole, PBS: polarizing beam splitter, BS1-2: beam splitter, QWP: quarter-wave plate, OL: objective lens. The inset images indicated by red dashed lines show, from left to right, the MLA focusing the incident plane wave onto the curved micromirror array, the measured CGH point, and the SLM phase image}
    \label{fig:Experimental setup}
\end{figure}

We have our custom MLA design printed using two-photon lithography (by Printoptix GmbH) with \SI{70}{\micro\meter} pitch and \SI{1}{\milli\meter} focal length, then built an experimental setup to validate our MLA correction scheme (Fig.~\ref{fig:Experimental setup}). The SLM surface is imaged with digital holographic microscopy (DHM), which allows us to characterize the surface phase profile and verify the effectiveness of the correction. Our setup also provides simultaneous measurement of the CGH-generated far-field point cloud to quantify optical performance. The light source is a 532 nm diode-pumped solid-state laser (Thorlabs CPS532), whose beam passes through a spatial filter composed of an L1 (10$\times$ objective) and L2 (f=\SI{180}{\milli\meter}) lens pair and a pinhole (\SI{25}{\micro\meter}) to remove high-frequency noise. The filtered beam was then split into a sample beam and a reference beam using a polarizing beam splitter (PBS). The sample beam was directed through a relay composed of an L3 (f=\SI{200}{\milli\meter}) and OL  (2$\times$ objective) lens pair to illuminate the SLM with a plane wave. The SLM is mounted on a 5-axis translation and tilt stage to enable precise positioning and selection of the measurement area. The reflected light from the SLM was collected through the same objective lens and then split by a second beam splitter (BS2) into two separate paths, each with a separate camera sensor (FLIR Blackfly 3.0). One camera (CGH CAM) was located after L4 (f=\SI{100}{\milli\meter}) at the Fourier plane to monitor the far-field holographic focus, while the other (DHM CAM) was positioned at the conjugate plane of the SLM. The reference beam interferes with the object beam in this light path to record amplitude and phase profiles via DHM. 
The reference beam included an adjustable delay line to match the optical path length to that of the sample beam, enabling coherent interference at the DHM camera. Interference images were processed using the Hilbert transform to extract both amplitude and phase information from the SLM surface~\cite{hilberttransform}. We added a quarter-wave plate (QWP) in the optical path to suppress unwanted back reflections from the other optical components. Finally, we inserted the MLA between the SLM and the objective lens; further details on the alignment and procedures can be found in the Supplementary Document.

\subsection{Validating phase flattening by the microlens array}
\begin{figure}[h]
    \centering
    \includegraphics[width=0.8\textwidth]{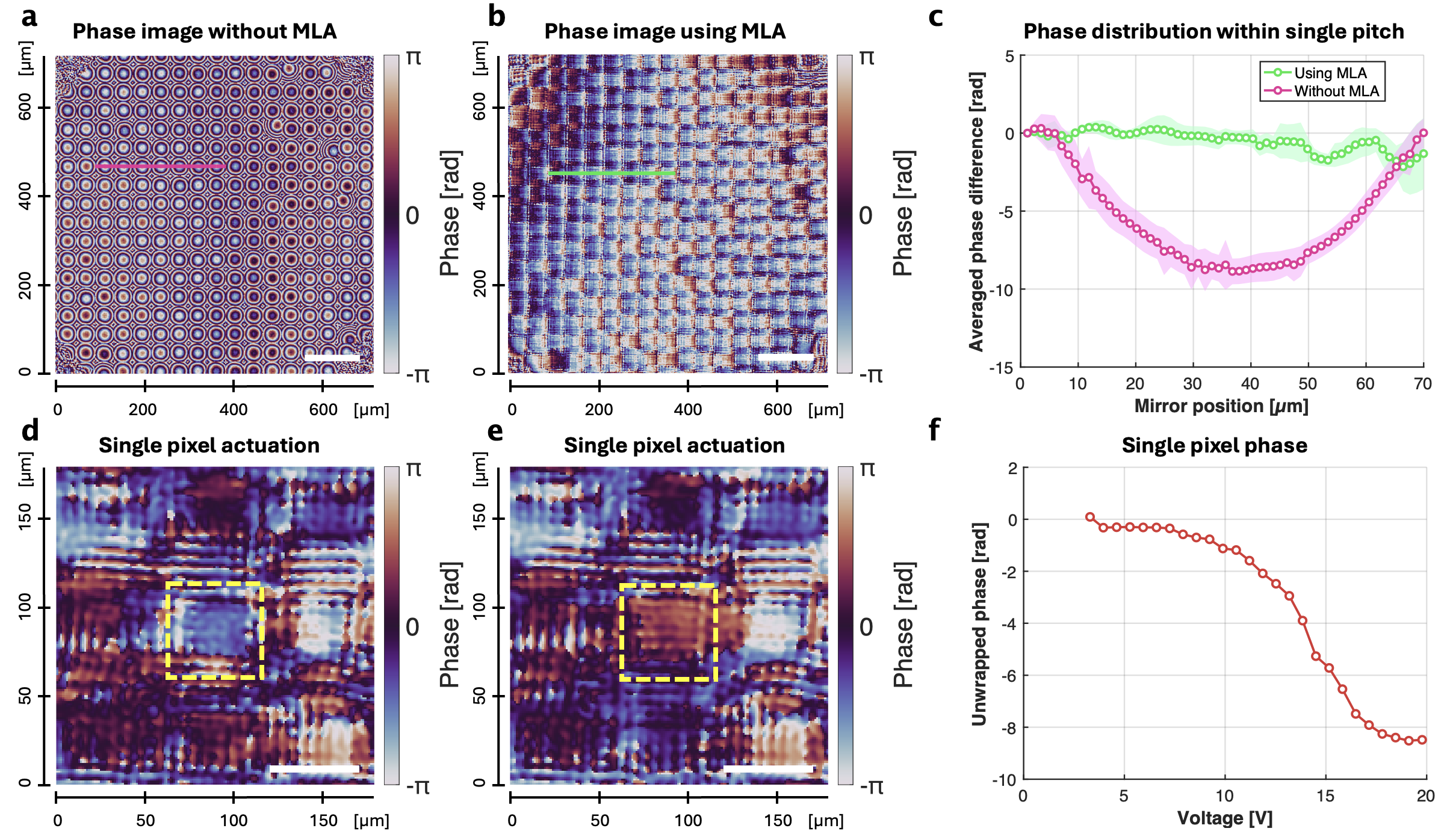}
    \caption{Experimental phase measurement of the SLM using DHM. (a) Measured phase image of the SLM before flattening. (b) Phase image after flattening using the MLA. (c) Phase distribution within a single pitch, obtained by averaging the phase values of individual mirrors along the pink line in (a) and the light green line in (b). (d, e) Phase response of the SLM to single pixel actuation after MLA-based flattening. A voltage was applied only to the mirror enclosed by the yellow dashed square. (f) Phase modulation curve of a single pixel under continuous voltage variation. The colorbar indicates phase in radians. Scalebars: \SI{200}{\micro\meter} for (a, b), and \SI{50}{\micro\meter} for (d, e).}
    \label{fig:Experimental results}
\end{figure}

Figure~\ref{fig:Experimental results} shows phase images of the SLM surface measured by DHM. As shown in Fig.~\ref{fig:Experimental results}a, the micromirrors exhibit a spherical phase profile across each mirror, a consequence of curvature. By unwrapping the phase along the line indicated in Fig.~\ref{fig:Experimental results}a and plotting the phase distribution within a single mirror pitch, we observe an average phase variation of approximately 8.6 radians, corresponding to a surface height difference of about 360~nm (Fig.~\ref{fig:Experimental results}c).
In contrast, Fig.~\ref{fig:Experimental results}b shows the phase image acquired after introducing the MLA. The overall curvature is almost flattened. The unwrapped phase profile along the indicated line in Fig.~\ref{fig:Experimental results}b demonstrates that the phase remains nearly flat across each mirror pitch (Fig.~\ref{fig:Experimental results}c), confirming that the MLA effectively avoids the inherent mirror curvature. From Fig.~\ref{fig:Experimental results}c, the radius of curvature of the mirror’s phase distribution is 1.7 mm before adding the MLA. After MLA correction, however, it is increased to 24.6 mm, representing an improvement by a factor of 15.5$\times$.

To evaluate dynamic control capability, we applied a voltage to a single micromirror and monitored the resulting phase change. As shown in Fig.~\ref{fig:Experimental results}d and e, a localized phase shift is clearly observed. A full voltage sweep confirmed that this pixel is capable of modulating the phase over a range exceeding $2\pi$ (Fig.~\ref{fig:Experimental results}f).
These results experimentally validate that the MLA can correct curvature-induced phase errors. This demonstrates the feasibility of our approach for real-world adaptive optics and dynamic wavefront shaping applications.

\subsection{Performance evaluation with Computer Generated Holography}

\begin{figure}[h]
    \centering
    \includegraphics[width=1\textwidth]{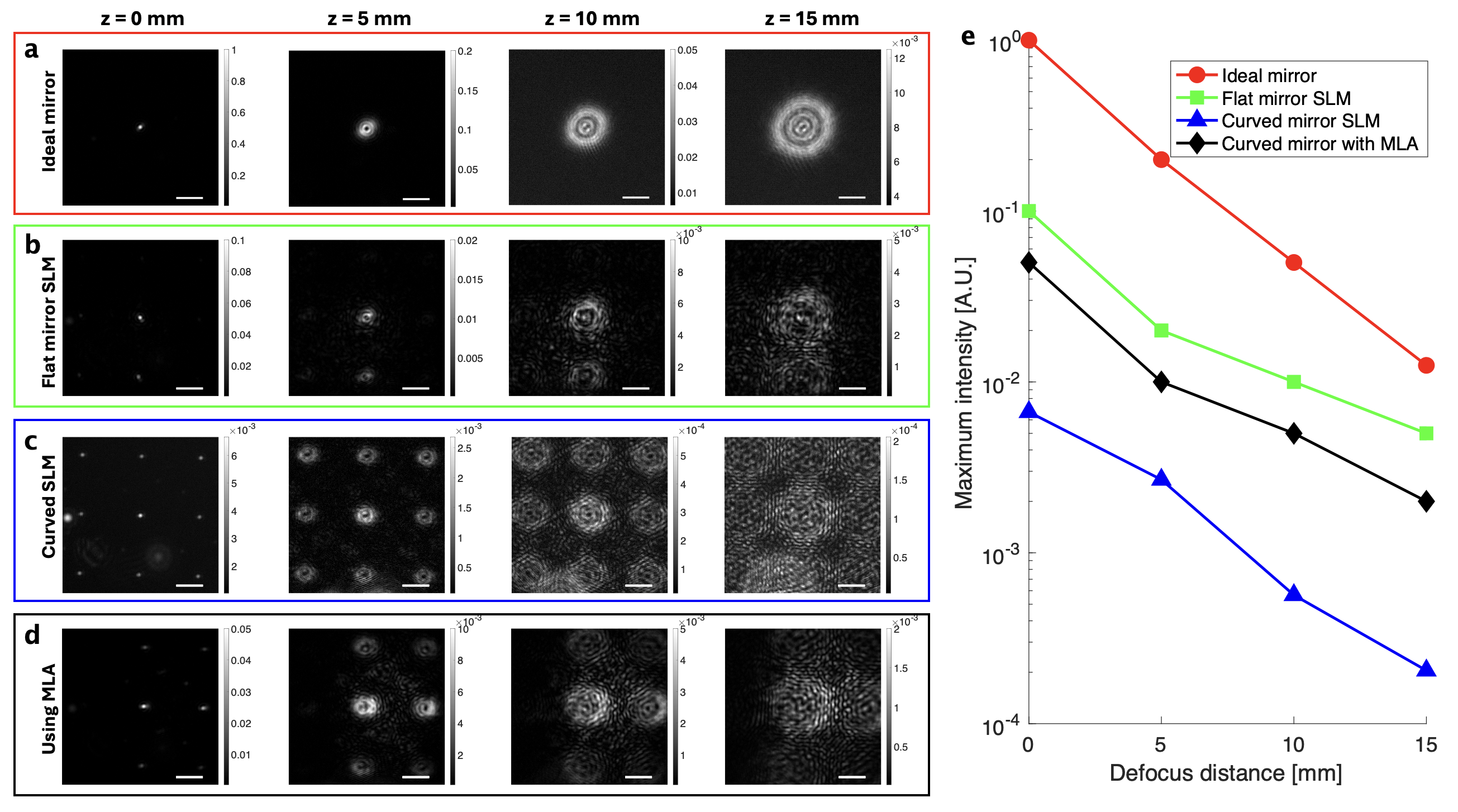}
    \caption{CGH point measurement for evaluating the optical performance of the SLM. To compare different mirror configurations, we measured the CGH-generated focal spot using an ideal mirror, a flat mirror SLM, a curved mirror SLM, and an MLA-flattened SLM. (a) Defocus response of the CGH spot generated by an ideal mirror, measured by physically translating the camera along the optical axis ($z = 0$, 5, 10, 15 mm from left to right). (b–d) Same as (a), but using a flat mirror SLM, a curved mirror SLM, and an MLA-corrected SLM, respectively. (e) Maximum intensity of the focal spot as a function of defocus distance for each case. The colorbar represents normalized intensity. Scalebars: \SI{200}{\micro\meter}.}
    \label{fig:Active phase control}
\end{figure}

After confirming the effective phase flattening with the MLA, we next investigated the impact of this correction on the generation of a single-spot CGH. We measured the CGH points produced by various types of mirrors in the sample arm using the CGH camera. The experimental conditions were matched to the simulation shown in Fig.~\ref{fig:SLM structure and simulation}, and we observed the light propagation in 3D by physically translating the CGH camera along the optical axis.
Figure~\ref{fig:Active phase control}a shows the propagation results when using an ideal flat mirror in place of the SLM. From left to right, the images correspond to propagation distances of 0 mm, 5 mm, 10 mm, and 15 mm, respectively. A well-defined focal spot is observed with gradual defocusing, as expected, providing a benchmark for SLM performance. For comparison, we also measured a separate SLM of the same geometry but which was fabricated with a much thicker polysilicon mirror body to reinforce against the residual stress of the gold coating and achieve flatter mirrors~\cite{OMN}~(Fig.~\ref{fig:Active phase control}b). This device produced a similar 3D intensity profile, but with significant artifacts at larger defocus distances due to the imperfect fill factor and diffraction from the pixels.
Figure~\ref{fig:Active phase control}c shows the result using the curved MEMS SLM without an MLA. As predicted by our simulation, strong aliasing artifacts are visible around the focal point, significantly degrading the intensity of the central spot.
When the MLA is introduced (Fig.~\ref{fig:Active phase control}d), the aliasing artifacts are substantially reduced, and the intensity of the focused point increases. Figure~\ref{fig:Active phase control}e shows the maximum intensity as a function of defocus distance for each case. The phase errors introduced by the curved mirror cause strong aliasing artifacts around the focal point, significantly degrading the intensity by a factor of approximately 160$\times$ compared to the ideal mirror. However, with the correction of the phase by the MLA, the peak intensity improves by about 8 times, reaching a similar order of magnitude as that observed with the flat mirror SLM.
These results demonstrate that MLA-based correction not only restores phase uniformity on curved mirror SLM surfaces but also enhances the quality and brightness of CGH points. This confirms the practical viability of our method for holographic projection using MEMS-based SLMs with curved mirrors.

\section{Discussion and Conclusion}
In summary, we demonstrate that the addition of a pitch-matched MLA to a micromirror-based SLM corrects for curvature-induced phase distortions and effectively decouples optical fill factor from mechanical fill factor. We can thus improve the generation of holographic points without adding costly or complicated steps to the micromirror array fabrication process.

There remain opportunities for further improvement and optimization in future work.

The proposed approach offers powerful capabilities for wavefront correction, provided that the alignment between the MLA and SLM is carefully maintained. Small translational or rotational offsets can affect overall performance, and high spatial frequency components may introduce optical crosstalk between adjacent microlenses. Additionally, the lenslet boundaries can produce phase discontinuities, particularly near mirror edges. These characteristics point to important design and integration opportunities. By refining MLA architectures to mitigate crosstalk and enhance tolerance to misalignment, future work can enable even greater robustness and effectiveness of MLA-based correction in demanding optical applications.

The proposed MLA-based correction method is generally applicable to micromirror-based SLMs that exhibit pixel-level curvature. The main requirements are that the MLA pitch matches the pixel pitch of the SLM, and that the MLA is positioned at the appropriate focal distance relative to the mirror array. Because this approach reshapes the incident beam, it does not require any modification to the SLM device itself, and can in principle be extended to other MEMS-based piston-motion SLM architectures.

For commercially available SLMs, however, there are additional practical considerations. Many existing devices have relatively small pixel pitches (e.g., < \SI{10}{\micro\meter}) and include protective cover layers, which can make it more demanding to achieve sufficiently small focal spots using an external MLA. Nevertheless, for emerging high-speed MEMS-based SLM designs, particularly those with larger pixel sizes or co-designed optical interfaces, the proposed approach offers a simple and scalable correction strategy.

While our results here focus on using MLAs for curvature correction, incorporating an MLA into the design process can also enable co-designed low fill factor MMAs ~\cite{cowan98MLA, CMOSMLA} that are easier to fabricate (improving cost and yield) and achieve better system performance. For the \SI{70}{\micro\meter}-pitch system presented, our approach uses $\leq 10\%$ of the original active mirror area while achieving an optical fill factor approaching 100\%. Future high-speed micromirror arrays designs for holography can thus use even smaller mirrors with more space between mirrors (i.e. lower mechanical fill factor) without compromising optical fill factor. Smaller mirrors are advantageous for maximizing SLM refresh rates because they can achieve higher resonant frequencies and faster step responses due to their lower mass and reduced squeeze-film damping. Designs with lower mechanical fill factor are also more robust to lithography errors and etch process variations, with the overall effect of improving yield. Codesigning an MLA and a piston-motion MMA-based SLM thus offers a scalable path towards high-speed computer generated holography for VR/AR, holographic optogenetics, and 3D fabrication. Ultimately, our proposed approach bridges the gap between the performance of LCoS-based SLMs and the high-speed actuation of MEMS-based devices.

\section*{Funding}
This work was supported by National Institutes of Health Research Project R01 Grant No. 1RF1NS131075-01, U.S. Air Force Office of Scientific Research Award No. FA955-22-1-0521, and Weill Neurohub Investigators Program. Leyla A. Kabuli was supported by the National Science Foundation Graduate Research Fellowship Program under Grant DGE 2146752. Elizabeth Murray was supported by the Apple Fellowship in Integrated Systems. Laura Waller is a Chan Zuckerberg Biohub SF investigator. 

\section*{Acknowledgments}
The authors would like to thank Nathan Tessema Ersaro and Cem Yalcin for helpful discussions. Leyla A. Kabuli was supported by the National Science Foundation Graduate Research Fellowship Program under Grant DGE 2146752.  Laura Waller is a Chan Zuckerberg Biohub SF investigator.

\section*{Disclosures}
The authors declare no conflicts of interest.

\section*{Supplemental document}
See Supplementary Document for additional implementation details and results.

\newpage

\renewcommand{\thesection}{S\arabic{section}}
\renewcommand{\thetable}{S\arabic{table}}
\renewcommand{\thefigure}{S\arabic{figure}}
\renewcommand{\theHfigure}{S\arabic{figure}}
\renewcommand\theequation{S\arabic{equation}}
\renewcommand{\theHequation}{S\arabic{equation}}
\renewcommand{\theHsection}{S\arabic{section}}
\renewcommand{\theHsubsection}{S\arabic{section}.\arabic{subsection}}
\setcounter{figure}{0}
\setcounter{table}{0}
\setcounter{section}{0}
\setcounter{equation}{0}

\noindent \textbf{\large Microlens Array-Based Optical Phase Flattening of Fabrication-Induced Curvature in Micromirror Devices: Supplementary Document}

\section{Angular Spectrum Propagation for Curved Mirror SLMs}

To model the propagation of light through our optical system, we employed the angular spectrum method~\cite{goodman2005fourieroptics}. In our simulation, we start with a plane wave incident on the curved mirror array of our SLM at the SLM plane with spatial coordinates \((x,y)\). The reflected field from this curved mirror array then propagates through a lens that focuses the light to a focal plane.

First, we model the incident plane wave at the SLM plane as
\begin{equation}
E_{in}(x,y) = A_0 e^{i\phi_0},
\end{equation}
where \(A_0\) is the amplitude and \(\phi_0\) is the initial phase of the plane wave.

The phase profile introduced by the curved mirror array can be modeled as
\begin{equation}
\phi_{m}(x,y) = \frac{k}{2R}(x^2 + y^2),
\end{equation}
where \(R\) is the radius of curvature of the mirror, \(k = 2\pi/\lambda\) is the wave number and \(\lambda\) is the wavelength.

The field reflected from the curved mirror array, which serves as our initial field for propagation analysis, is given by
\begin{equation}
E_{SLM}(x,y) = E_{in}(x,y) \cdot e^{i2\phi_{m}(x,y)}.
\end{equation}
The factor of 2 in the exponent accounts for the round-trip phase change upon reflection.

This reflected field then propagates to a lens with a focal length \(f\) and is focused in the focal plane. In the focal plane, the field can be expressed in terms of spatial coordinates \((x',y')\):

\begin{equation}
E_{focal}(x', y') \approx \frac{e^{ikf}}{i\lambda f} e^{i  \frac{k}{2f}(k_x^2+k_y^2)} \mathcal{F}\{E_{SLM}(x,y)\}\left(\frac{x'}{\lambda f}, \frac{y'}{\lambda f}\right)
\end{equation}
where \(\mathcal{F}\) represents the Fourier transform. The spatial coordinates \((x',y')\) in the focal plane are related to the spatial frequencies \((k_x,k_y)\) by the relations \(k_x = \frac{x'}{\lambda f}\) and \(k_y = \frac{y'}{\lambda f}\).

To analyze the effect of the focused spot, we propagated the field from the focal plane by varying the propagation distance \(z\) using the angular spectrum method:
\begin{equation}
E(x',y';z) = \mathcal{F}^{-1}\left\{\mathcal{F}\{E_{focal}(x',y')\} \cdot H(k_x, k_y; z)\right\}
\end{equation}
where $\mathcal{F}$ represents the inverse Fourier transform, and the transfer function \(H(k_x, k_y; z)\) is given by:
\begin{equation}
H(k_x, k_y; z) = e^{iz\sqrt{k^2 - k_x^2 - k_y^2}}
\end{equation}

The results indicated that the mirror's curvature would cause significant distortion in wavefront shaping.

\section{Design and Fabrication of a Piston-Motion Micromirror-Based SLM}
\begin{figure}[h]
    \centering
    \includegraphics[width=0.8\linewidth]{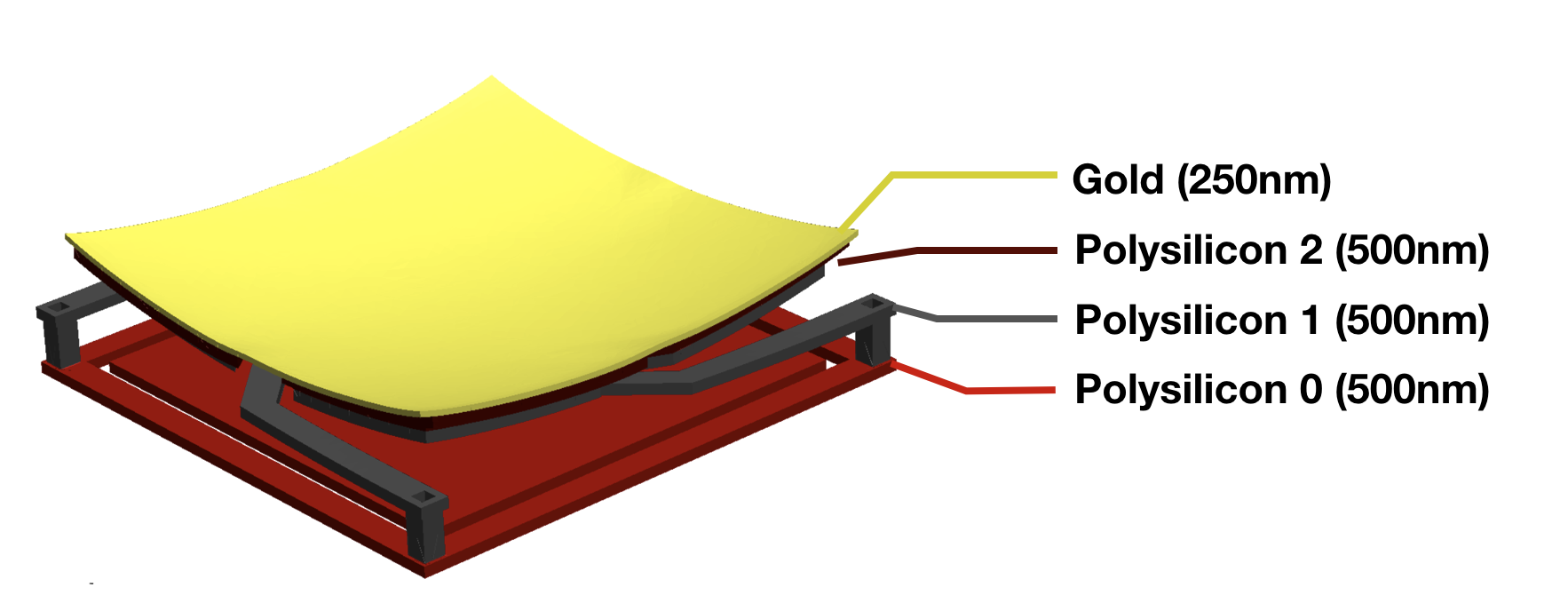}
    \caption{Layer stackup and geometry of a single mirror in the \SI{70}{\micro\meter}-pitch micromirror array.}
    \label{fig:mirror}
\end{figure}

The \SI{70}{\micro\meter}-pitch mircromirror array was designed and fabricated in a modified Poly-MUMPS surface micromachining process. The structure of an individual mirror pixel is shown in Fig. ~\ref{fig:mirror}. Each mirror is actuated electrostatically and consists of three LPCVD polysilicon structural layers and an evaporated gold layer for reflectivity. The high residual stress of the gold layer deposited on a thin polysilicon mirror body results in unwanted mirror curvature after release. 

\section{Alignment of MLA and Effectiveness Verfication}
Precise alignment between MLA and SLM is critical for effective phase flattening. Since the MLA should focus the light to the central region of each curved micromirror, any lateral misalignment or incorrect axial spacing between the MLA and SLM can significantly degrade performance. In this section, we describe our approach to achieve accurate alignment of the MLA using transmission imaging.

\begin{figure}[h]
    \centering
    \includegraphics[width=1\textwidth]{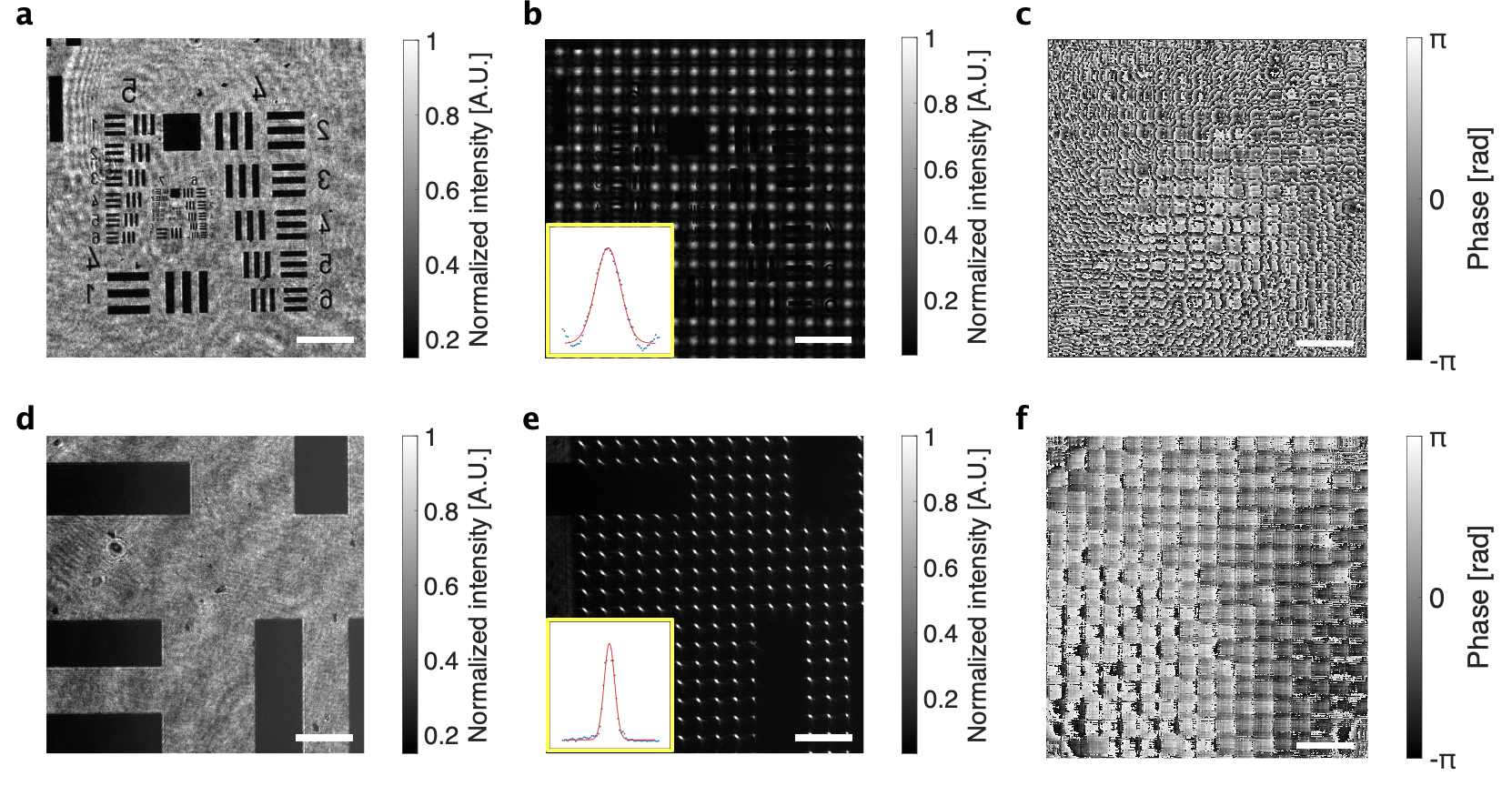}
    \caption{Effectiveness of MLA. (a) Transmission image of resolution target without the MLA. (b) Same in (a), but using MLA of \SI{3.2}{\milli\meter} focal length and \SI{72}{\micro\meter} pitch. The MLA generates array of focused spots. Inset image enclosed by yellow square is a PSF of a single spot. (c) Phase image with MLA with mismatch of the pitch between SLM and MLA. (d-f) Same as (a-c), but using different MLA with \SI{1}{\milli\meter} and \SI{70}{\micro\meter} The colorbars in (a), (b), (d), and (e) represent intensity. The colorbars in (c) and (f) indicate phase in radian. Scalebars: \SI{200}{\micro\meter}.}
    \label{fig:transmission}
\end{figure}

Previously, we placed the SLM at the right place by directly imaging the SLM surface in real-time using DHM. However, when we introduced the MLA, no longer have the direct image and there is an uncertainty of the MLA position even though we know the focal length of it. To address this, we built a transmission imaging path, temporarily replacing the SLM with a transparent USAF resolution target for direct optical inspection. By simultaneously monitoring the resolution target with DHM from the reflection path and the transmission path, we precisely determined the position of the image plane. Then, the MLA was added. Here, two different MLAs were tested. One with \SI{3.2}{\milli\meter} focal length and \SI{72}{\micro\meter} pitch (APO-Q-P72-R1.45, Advanced Microoptic Systems GmbH), and a custom MLA with \SI{1.0}{\milli\meter} focal length and \SI{70}{\micro\meter} pitch (Printoptix). Fig.~\ref{fig:transmission}a shows groups 4 and 5 of the resolution target. When inserting the \SI{3.2}{\milli\meter} focal length MLA, a focused spot array as in Fig.~\ref{fig:transmission}b was observed. The inset shows the point spread function (PSF) of a single spot with full width at half maximum (FWHM) of \SI{19.3}{\micro\meter}. Similarly, Fig.~\ref{fig:transmission}d shows elements 3 and 4 of group 1. With the \SI{1.0}{\milli\meter} focal length MLA, the focused spots (Fig.~\ref{fig:transmission}e) were significantly smaller, with a FWHM of \SI{8.2}{\micro\meter}. After MLA position was fixed, the resolution target was replaced with the SLM, and DHM images were used to fine-tune the lateral position of the MLA and the SLM.

\section{Effects of Pitch Mismatch Between MLA and SLM}
The first tested MLA had a 72 µm pitch, slightly larger than the 70 µm pitch of our fabricated SLM. As shown in Fig.~\ref{fig:transmission}c, phase flattening was partially achieved in the well-aligned central part of the field of view (FOV). However, at the edge of the FOV, focused spots are reflected by slightly off-center regions of the mirrors. This misalignment led to effects similar to reflections from tilted mirrors, producing phase ramps and distortions. In contrast, the second MLA with a matched 70 µm pitch achieved consistent phase flattening across the entire SLM surface (Fig.~\ref{fig:transmission}f). Accordingly, all subsequent experiments used the same pitch MLA for the correction.

\section{Effects of rotational mismatch between MLA and SLM}
\begin{figure}[h]
    \centering
    \includegraphics[width=0.8\textwidth]{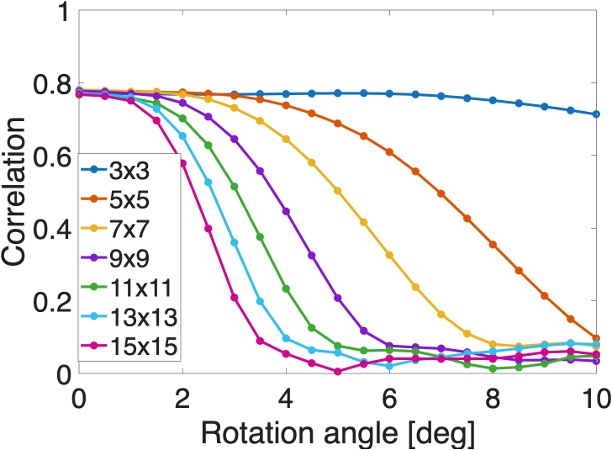}
    \caption{Complex field correlation with the ideal mirror as a function of the relative rotational angle between the micromirror array and the MLA, evaluated for different array sizes.}
    \label{fig:rotation}
\end{figure}
To evaluate the sensitivity of the system to rotational misalignment, additional simulations were performed in which a relative rotation between the mirror array and the MLA was introduced. (Fig.~\ref{fig:rotation})The results indicate that the tolerance to rotational mismatch decreases as the array size increases. For the smallest array size (3$\times$3), the complex field correlation remains above 95$\%$ until the relative rotation exceeds approximately \ang{8.6} with respect to the perfectly aligned case. In contrast, for a 15$\times$15 array, a rotation of about \ang{1.1} is sufficient to reduce the correlation below 95$\%$.

In our current device (N = 16), the micromirror array and MLA are aligned experimentally using a two-axis goniometric stage. Based on the stage specifications of the alignment stage, array sizes up to 32$\times$32 can be accommodated while maintaining a correlation above 95$\%$. The tolerable rotation therefore lies within a range that can be readily achieved using standard alignment procedures. With higher-precision alignment hardware, even larger array size are expected to be feasible.

\section{Simulation of MLA focal length dependence}
\begin{figure}[h!]
    \centering
    \includegraphics[width=0.8\textwidth]{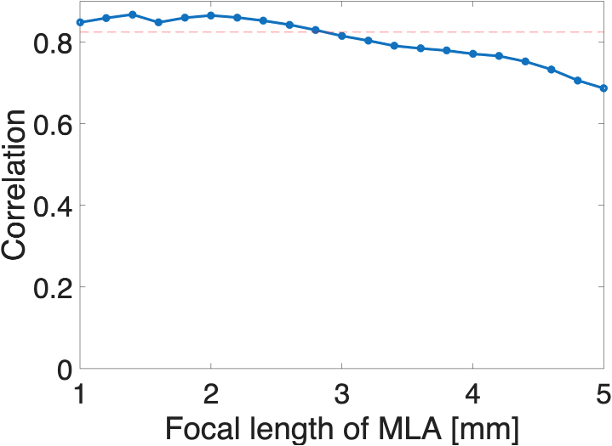}
    \caption{Complex field correlation with the ideal mirror as a function of the focal length of the MLA. The red dashed line indicates the 95$\%$ confidence level measured from the point of maximum correlation (1.4 mm focal length).}
    \label{fig:focal_length}
\end{figure}
For the curved mirror used in this work, the theoretically flat region has a diameter of approximately \SI{20.18}{\micro\meter}. To ensure robust performance, the optical configuration was designed to focus the point on a smaller diameter of \SI{9.27}{\micro\meter}. The dependence of the flattening performance on the focal length of the MLA was further investigated through additional simulations.(Fig.~\ref{fig:focal_length}) The results show that effective flattening is maintained for MLA focal lengths of up to approximately \SI{2.87}{\milli\meter}, corresponding to a spot diameter of \SI{26.61}{\micro\meter}. This behavior is expected to be generally applicable to other systems with similar configurations. Once the radius of curvature of the mirror and the operating wavelength are specified, the size of the flat region can be determined, and the MLA focal length can be selected accordingly. As a result, similar performance trends are anticipated for the systems with comparable parameters.

\section{Wavelength dependence of MLA-based Phase Flattening}
In practice, wavelength-dependent effects may arise from diffraction and material dispersion. The diffraction-limited focal spot size scales with wavelength, which can affect how the incident light is confined within the effective flat region of each micromirror. In addition, the wavelength dependence of the refractive index of the MLA material can introduce a slight change in the effective focal length, resulting in a small axial shift of the focal position. Furthermore, for piston-motion micromirrors, the phase modulation induced by a given physical displacement is inversely proportional to wavelength, which can also contribute to wavelength-dependent variations in the resulting field.

\begin{figure}[h!]
    \centering
    \includegraphics[width=1\textwidth]{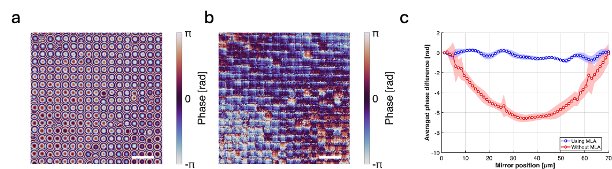}
    \caption{Experimental phase flattening at a wavelength of 635 nm. (a) Phase distribution before flattening. (b) Phase distribution after MLA-based flattening. (c) Averaged phase profile within a single mirror pitch. The red curve corresponds to the phase before flattening (a), and the blue curve corresponds to the flattened phase (b).}
    \label{fig:635nm}
\end{figure}

To evaluate these effects, we performed experimental and numerical analyses across different wavelength. In addition to the 532 nm wavelength used in the main manuscript, we experimentally verified the phase-flattening performance at 635 nm. At this wavelength, the micromirror exhibited an initial phase curvature of approximately 6.5 rad within a single pitch, which was effectively flattened to a nearly uniform phase distribution, comparable to the result obtained at 532 nm.

\begin{figure}[h!]
    \centering
    \includegraphics[width=0.8\textwidth]{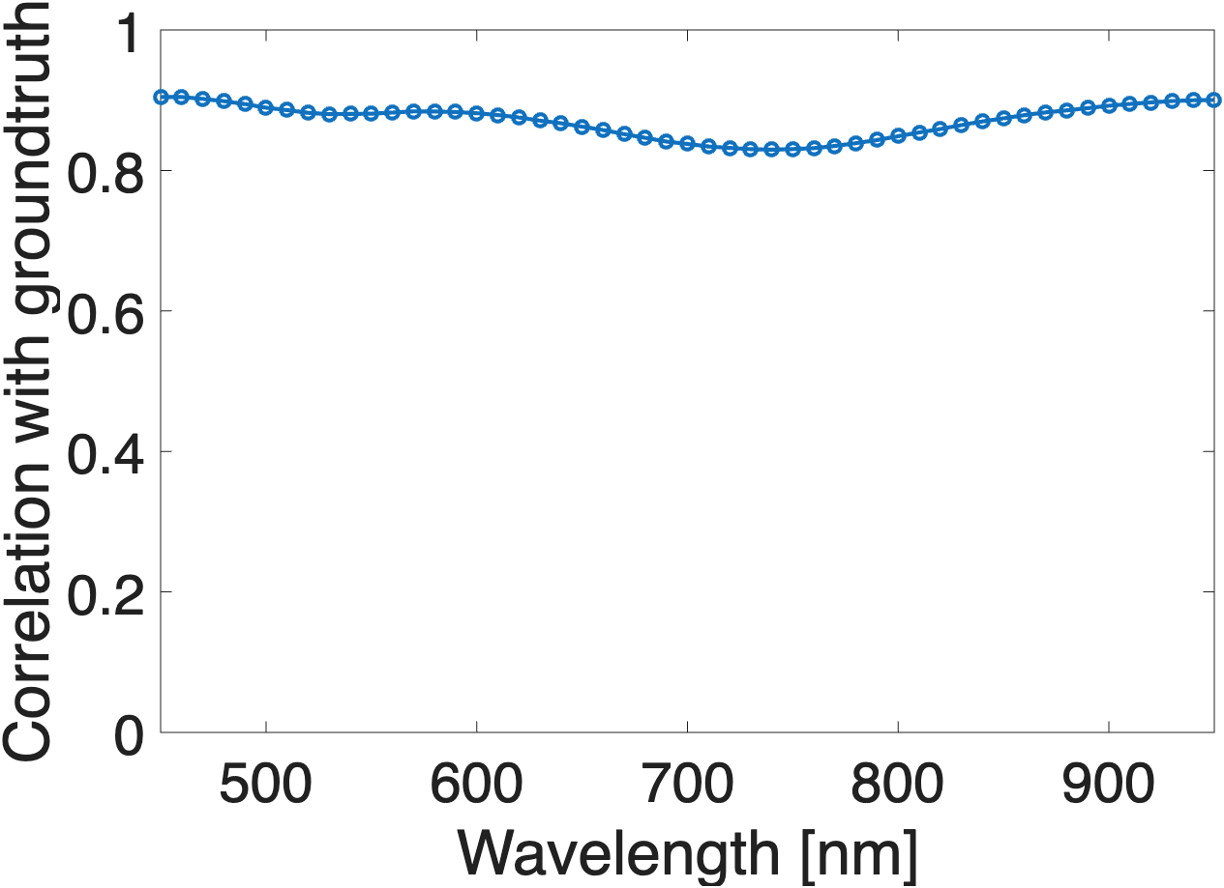}
    \caption{Complex field correlation as a function of wavelength (450–950 nm) for the MLA-based correction, showing consistently high correlation across the entire wavelength range.}
    \label{fig:wl_simul}
\end{figure}

We further conducted numerical simulations over a broad wavelength range (450–950 nm) and compared the complex field correlation between the MLA-corrected curved mirror and the corresponding MLA-corrected flat mirror. The results show that the correlation remains high across the entire wavelength range, with an average value of approximately 0.87 and a standard deviation of about 0.02.

These results indicate that, although diffraction and dispersion introduce wavelength-dependent effects, the overall flattening performance of the proposed MLA-based approach is not strongly dependent on wavelength over the considered range.

\end{document}